\documentclass[12pt]{article}
\usepackage{epsfig}

\def\beq{\begin{equation}}
\def\eeq#1{\label{#1}\end{equation}}
\def\eeqn{\end{equation}}

\def\beqa{\begin{eqnarray}}
\def\eeqa#1{\label{#1}\end{eqnarray}}
\def\eeqan{\end{eqnarray}}

\let\bar=\overbar

\def\Dslash{\not{\hbox{\kern-4pt $D$}}}
\def\dslash{\not{\hbox{\kern-2pt $\del$}}}

\def\msb{{\bar{\ssstyle M \kern -1pt S}}}

\def\Title#1{\begin{center} {\Large {\bf #1} } \end{center}}

\begin{document}

\Title{$\phi_{1}$ Measurements at Belle}

\bigskip\bigskip


\begin{raggedright}  

{\it Veronika Chobanova\index{Chobanova, V.}\\
Max-Planck-Institut f\"ur Teilchenphysik\\
D-80805 M\"unchen, Germany}
\bigskip
\bigskip
\end{raggedright}

\begin{center}
Presented at \\ \it{Flavor Physics and CP Violation (FPCP 2014) \\ Marseille, France, May 26-30, 2014}
\end{center}
\bigskip

We present the recent measurements of the $B^{0}\rightarrow\eta' K^{0}$ and the $B\rightarrow\omega K$ decay modes based on the full data set of $772\times10^6$
$B\bar{B}$ pairs collected by the Belle detector at the KEKB $e^{+}e^{-}$ collider. In the $B^{0}\rightarrow\eta' K^{0}$ mode, we obtain the $CP$-violating parameters
\begin{eqnarray}
  {\cal A}_{\eta'K^{0}} &=& +0.03\pm0.05\,{\rm(stat)}\pm0.04\,{\rm(syst)}, \nonumber \\
  {\cal S}_{\eta'K^{0}} &=& +0.68\pm0.07\,{\rm(stat)}\pm0.03\,{\rm(syst)}. \nonumber 
\end{eqnarray}
This is the world's most precise result on the $\eta' K^{0}$ $CP$ parameters. In $B\rightarrow\omega K$ decays, we measure the branching fractions 
\begin{eqnarray}
  {\cal B}(B^{0}\rightarrow\omega K^{0}) &=& (4.5\pm0.4\,{\rm(stat)}\pm0.3\,{\rm(syst))\times 10^{-6}}, \nonumber \\
  {\cal B}(B^{-}\rightarrow\omega K^{-}) &=& (6.8\pm0.4\,{\rm(stat)}\pm0.4\,{\rm(syst))\times 10^{-6}}, \nonumber
\end{eqnarray}
which are their current most precise results.
We measure the first evidence of $CP$ violation in the $B^{0}\rightarrow\omega K^{0}_{S}$ decay mode, obtaining the $CP$-violating parameters
\begin{eqnarray}
  {\cal A}_{\omega K^{0}_{S}} &=& -0.36\pm0.19\,{\rm(stat)}\pm0.05\,{\rm(syst)}, \nonumber \\
  {\cal S}_{\omega K^{0}_{S}} &=& +0.91\pm0.32\,{\rm(stat)}\pm0.05\,{\rm(syst)}. \nonumber 
\end{eqnarray}
In the $B^{-}\rightarrow\omega K^{-}$ mode, we measure the direct $CP$-violation parameter 
\begin{eqnarray}
  {\cal A}_{\omega K^{-}}= -0.03\pm0.04\,{\rm(stat)}\pm0.01\,{\rm(syst)}, \nonumber
\end{eqnarray}
which is its most precise measurement to date.
\section{Introduction}
The main purpose of the $B$ factories, Belle and BaBar, was to test the Cabibbo-Kobayashi-Maskawa (CKM) mechanism, which is a part of the Standard Model of particle 
physics (SM). The CKM mechanism foresees the existence of a $CP$-violating complex phase in the quark sector that gives rise to $CP$ violation. The $B$ factories
measured the angles, $\phi_{1}$, $\phi_{2}$ and $\phi_{3}$, and the sides of the unitarity triangle for $B$ mesons, the area of which is proportional to the SM 
$CP$-violating phase. In the past decade, measurements of the $\sin{2\phi_{1}}$ value in $b\rightarrow c \bar{c} s$ transitions confirmed the CKM mechanism predictions.
Despite the great success of the CKM theory, it fails to explain the magnitude of the matter-antimatter asymmetry in today's Universe. Thus, attention has turned 
towards search for new sources of $CP$ violation, introduced by new physics. Decays of the $B$ meson that proceed predominantly through loop-dominated 
$b\rightarrow s q \bar{q}$ transitions are highly sensitive to heavy particle contributions in the loop, which can be introduced by so far undetected particles. 
The decays presented in these proceedings, $B^{0}\rightarrow\eta'K^{0}$ and $B\rightarrow\omega K$, fall into this category. They are both sensitive to $\phi_{1}$, 
which can be accessed by measuring their time-dependent $CP$ asymmetries.

The $B$ factories operate at the $\Upsilon(4S)$ resonance, which decays almost exclusively into a $B\bar{B}$ pair. For flavour tagging of a $B$ meson of interest, we use 
flavour-specific decays of the other $B$ meson in the pair, the so-called ``tag side''. The time-dependent $CP$ asymmetry for a $B$ decay into a given $CP$ final state $f$, 
is given by
\begin{eqnarray}
 a_{CP}(\Delta t) = {\cal A}_f \cos{\Delta m_{d}\Delta t}+{\cal S}_f \sin{\Delta m_{d}\Delta t}, \nonumber
\end{eqnarray}
where $\Delta t$ is the difference in the decay time of the two $B$ mesons produced in the $\Upsilon(4S)$ decay, $\Delta m_{d}$ is the difference between the masses of 
the heavy and light mass eigenstates of the $B^{0}$ meson, and ${\cal A}_f$ and ${\cal S}_f$ are the $CP$-violating parameters. Assuming only a loop contribution in 
$\eta'K^{0}$ and in $\omega K^{0}_{S}$, we expect ${\cal S}_{f}=-\xi_{f}\sin{2\phi_{1}}$, where $\xi_{f}=\pm1$ is the $CP$ eigenvalue of the final state.
Due to pollution by lower-order processes, the measured ${\cal S}_{f}$ value deviates from this expectation by an amount 
$\Delta {\cal S}_{f} = {\cal S}_{f} - \sin{2\phi_{1}}$ that depends on the decay channel .

We present the measurements of the $B^{0}\rightarrow\eta' K^{0}$ and $B\rightarrow\omega K$ $CP$-asymmetry parameters, along with a measurement of the 
$B\rightarrow\omega K$ branching fractions, based on the full Belle~\cite{BelleDetector} data set containing 772 millions $B\bar{B}$ pairs.

\section{Measurement of $CP$ violation parameters in the $B^{0}\rightarrow\eta' K^{0}$ decay}
In the $B^{0}\rightarrow\eta' K^{0}$ decay channel, the Standard Model predicts $-0.05 \leq \Delta {\cal S}_{\eta'K^{0}} \leq 0.09$~\cite{etapk_theory}. 
The $CP$ parameters of the two $CP$ final states $\eta' K^{0}_{S}$ and $\eta' K^{0}_{L}$ are measured separately and in the end combined to a common value, since they are 
equal according to the theory.

\begin{figure}   
  \includegraphics[width=0.59\textwidth,bb=80 400 500 530,clip=true]{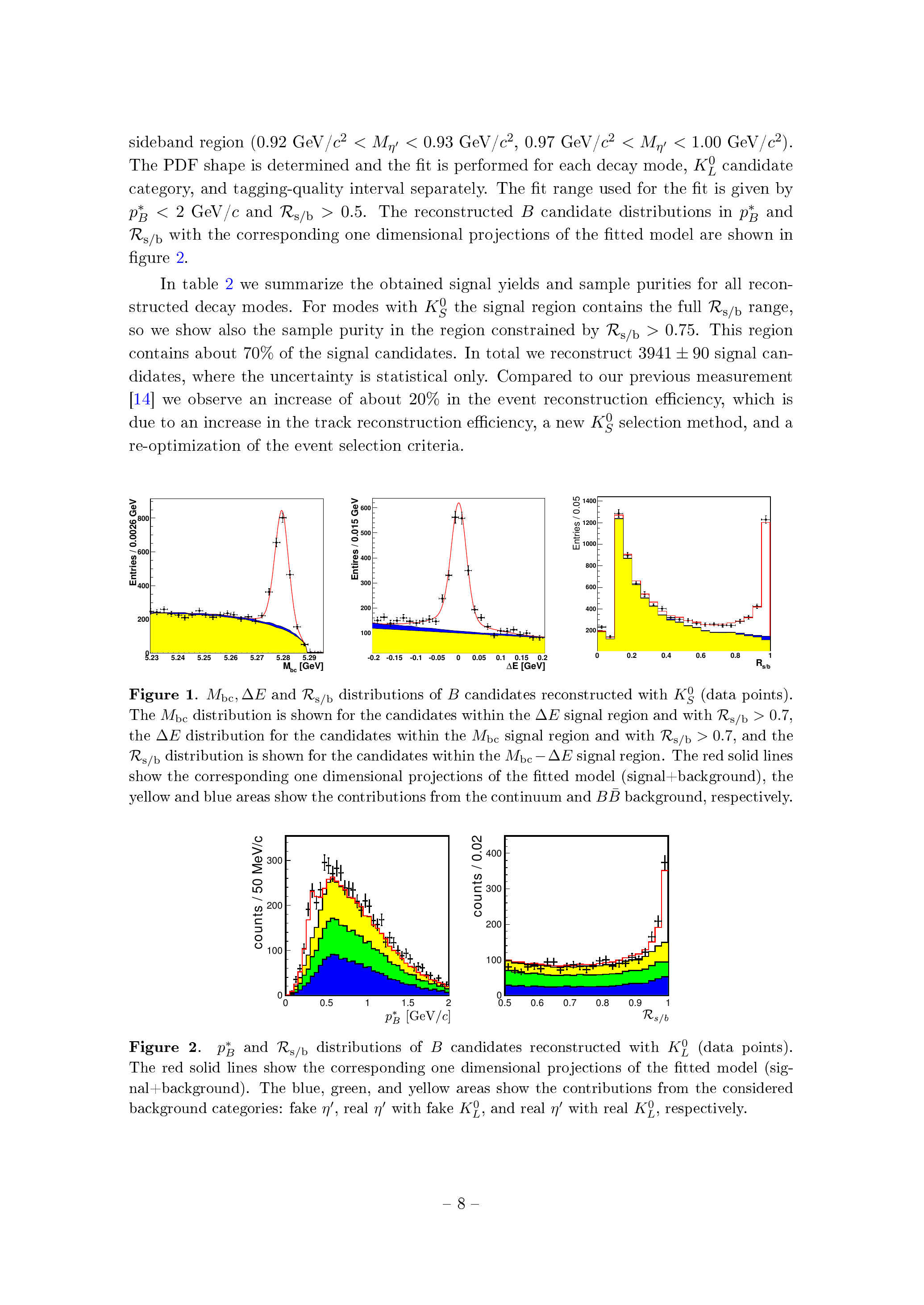}
  \includegraphics[width=0.39\textwidth,bb=160 180 435 310,clip=true]{b2etapk_fit.pdf}
  \caption{Projections of the fit to the data for the fit observables of $\eta'K^{0}_{S}$ (first three plots from the right) and $\eta'K^{0}_{L}$ (fourth and fifth plot).
   For $\eta'K^{0}_{S}$, the total PDF is given in red, the $q\bar{q}$ background contribution - in yellow and the $B\bar{B}$ background contribution - in blue.
   For $\eta'K^{0}_{L}$, the total PDF is in red, the combinatorial backgrounds with fake $\eta'$, with fake $K_{L}^{0}$ and with both real $\eta'$ and $K_{L}^{0}$ are given 
   in blue, green and yellow, respectively.}
  \label{etapk_fit}
\end{figure}

\begin{figure}
  \begin{center}
    \includegraphics[width=0.6\textwidth,bb=100 550 480 760,clip=true]{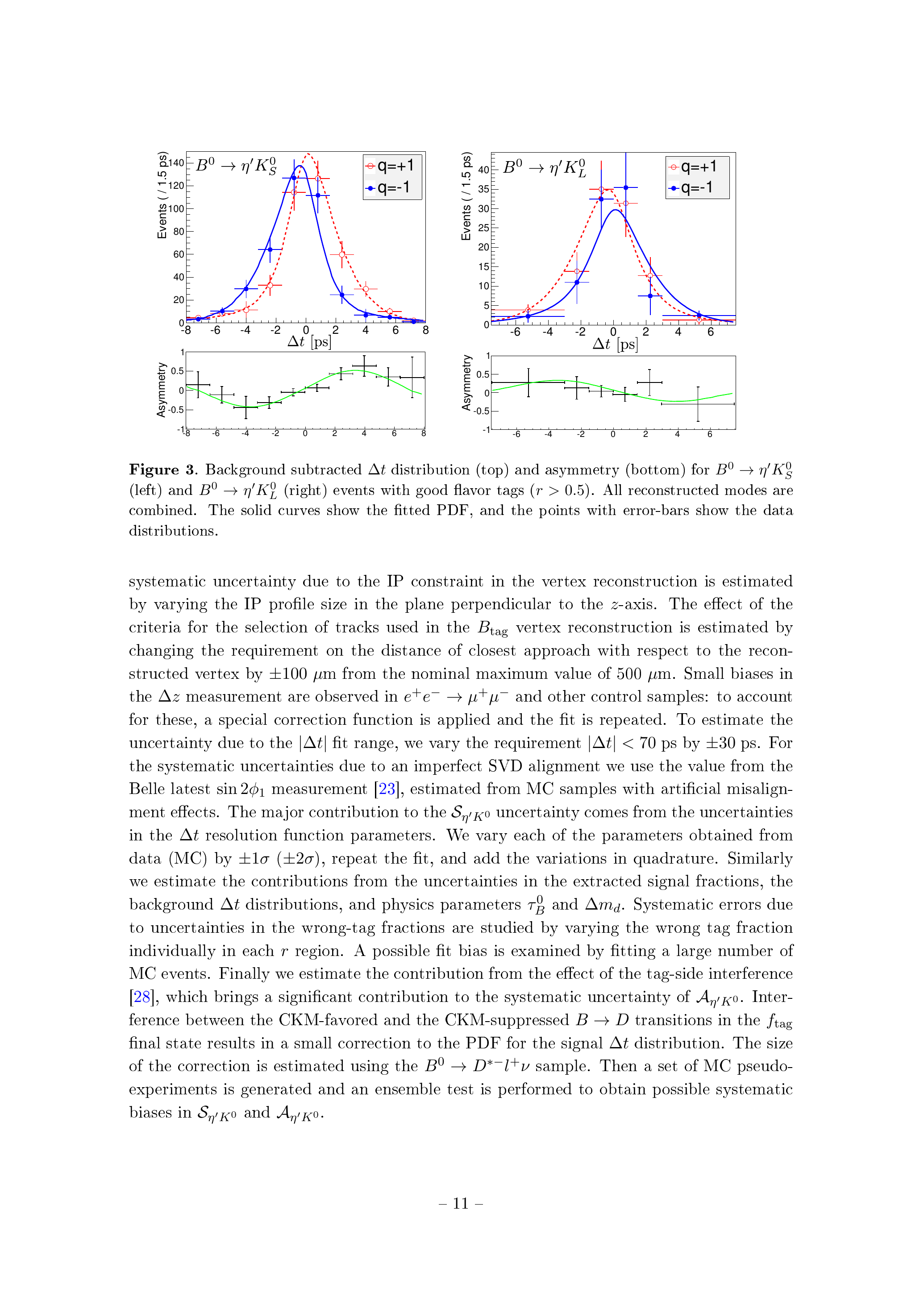}   
  \end{center}
  \caption{Projections of the data fit to the $\Delta t$ distribution for $\eta'K^{0}_{S}$ (left) and $\eta'K^{0}_{L}$ (right).}
  \label{etapk_cp}
\end{figure}

After the event reconstruction, we estimate the signal yield. For this, in the $\eta' K^{0}_{S}$ decay mode, we fit the event distributions of the beam-constrained mass
of the $B$ meson $M_{bc}$, the difference between the reconstructed $B$ meson energy and the beam energy $\Delta E$ and a $q\bar{q}$-background suppression likelihood ratio
$R_{s/b}$, based on event-shape variables. In the $\eta' K^{0}_{L}$ decay mode, we extract the signal yield by fitting $R_{s/b}$ and the $B$ meson momentum, reconstructed 
with the knowledge of the beam energy and the nominal $K^{0}_{L}$ mass. In total, we reconstruct $2503\pm63$ $\eta'K^{0}_{S}$ events and $1041\pm41$ $\eta'K^{0}_{L}$
signal events, where the uncertainties are statistical only. Following this, we fit the $\Delta t$ distribution for the two $B$ meson flavours $q$ to extract the $CP$
parameters, where $q=+1$ or $q=-1$ if the $B$ on the tag side is $B$ or $\bar{B}$. Projections of the fit to the data are shown in Fig.~\ref{etapk_cp}. Our preliminary results are
\begin{eqnarray}
{\cal A}_{\eta'K^{0}} &=& +0.03\pm0.05\,{\rm(stat)}\pm0.04\,{\rm(syst)}, \nonumber \\
{\cal S}_{\eta'K^{0}} &=& +0.68\pm0.07\,{\rm(stat)}\pm0.03\,{\rm(syst)}. \nonumber 
\end{eqnarray}
The main contribution to the $A_{\eta'K^{0}}$ systematic uncertainty comes from the tag-side interference effect and for $S_{\eta'K^{0}}$ - from the 
uncertainties in the $\Delta t$ resolution function parameters. Signal enhanced projections from the fit to the data are shown in Fig.~\ref{etapk_fit}. The measured values 
of $A_{\eta'K^{0}}$ and $S_{\eta'K^{0}}$ are the world's most precise values of $CP$ violation parameters in this particular decay and also out of all 
$b\rightarrow s q \bar{q}$ transition dominated decays. They are consistent with previous measurements by Belle~\cite{etapk_belle_prev} and BaBar~\cite{etapk_babar_prev} 
and with the SM predictions.

\section{Measurement of branching fractions and $CP$ violation parameters in $B\rightarrow\omega K$ decays}
We present the results from the measurements of the neutral decay mode $B^{0}\rightarrow \omega K^{0}_{S}$ and the charged decay mode $B^{-}\rightarrow \omega K^{-}$. 
The SM predicts $0.1 \leq \Delta {\cal S}_{\omega'K_{S}^{0}} \leq 0.2$~\cite{wk_theory_1,wk_theory_2,wk_theory_3,wk_theory_4}.

\begin{figure}   
  \includegraphics[width=0.3\textwidth]{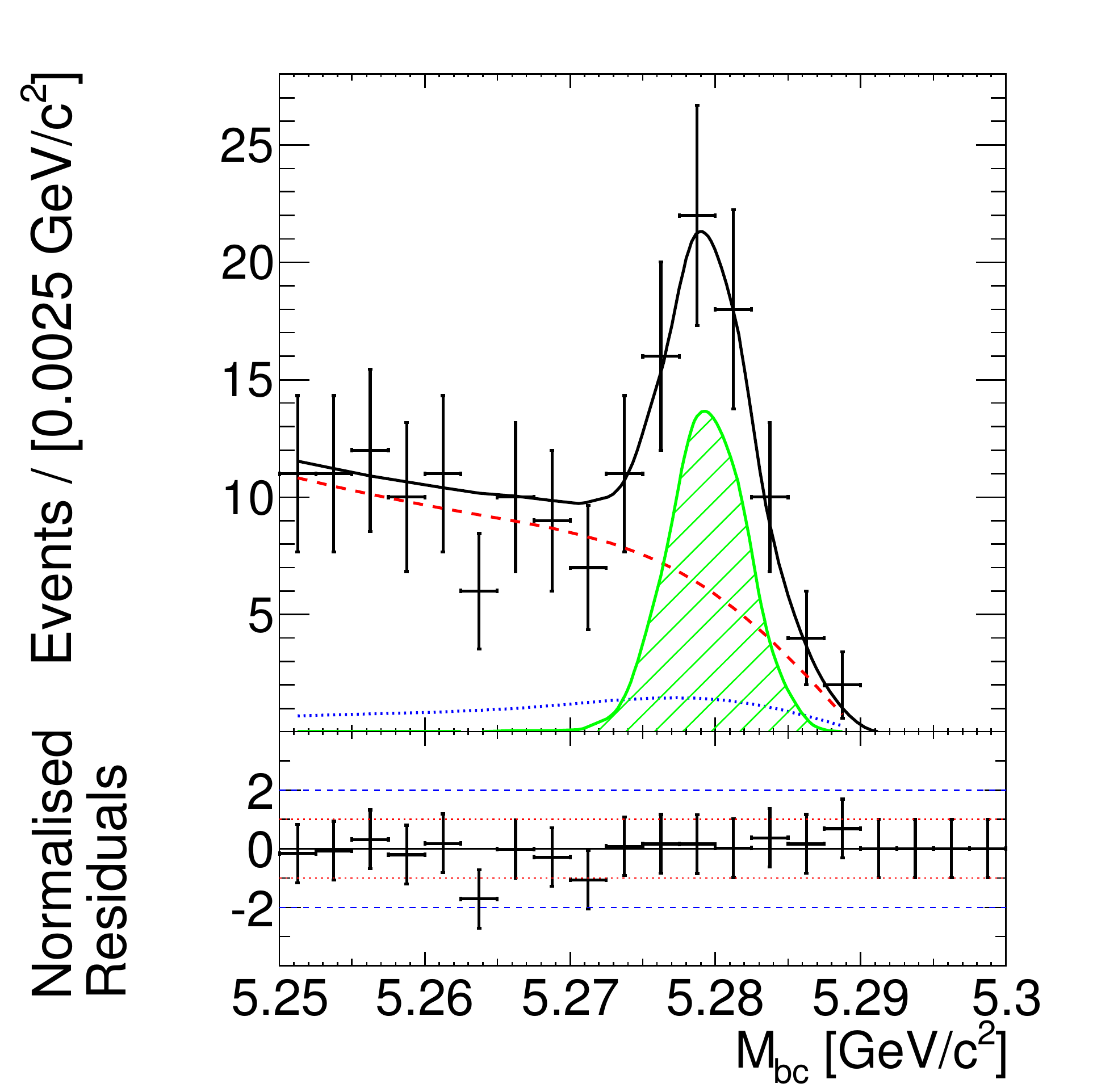}
  \includegraphics[width=0.3\textwidth]{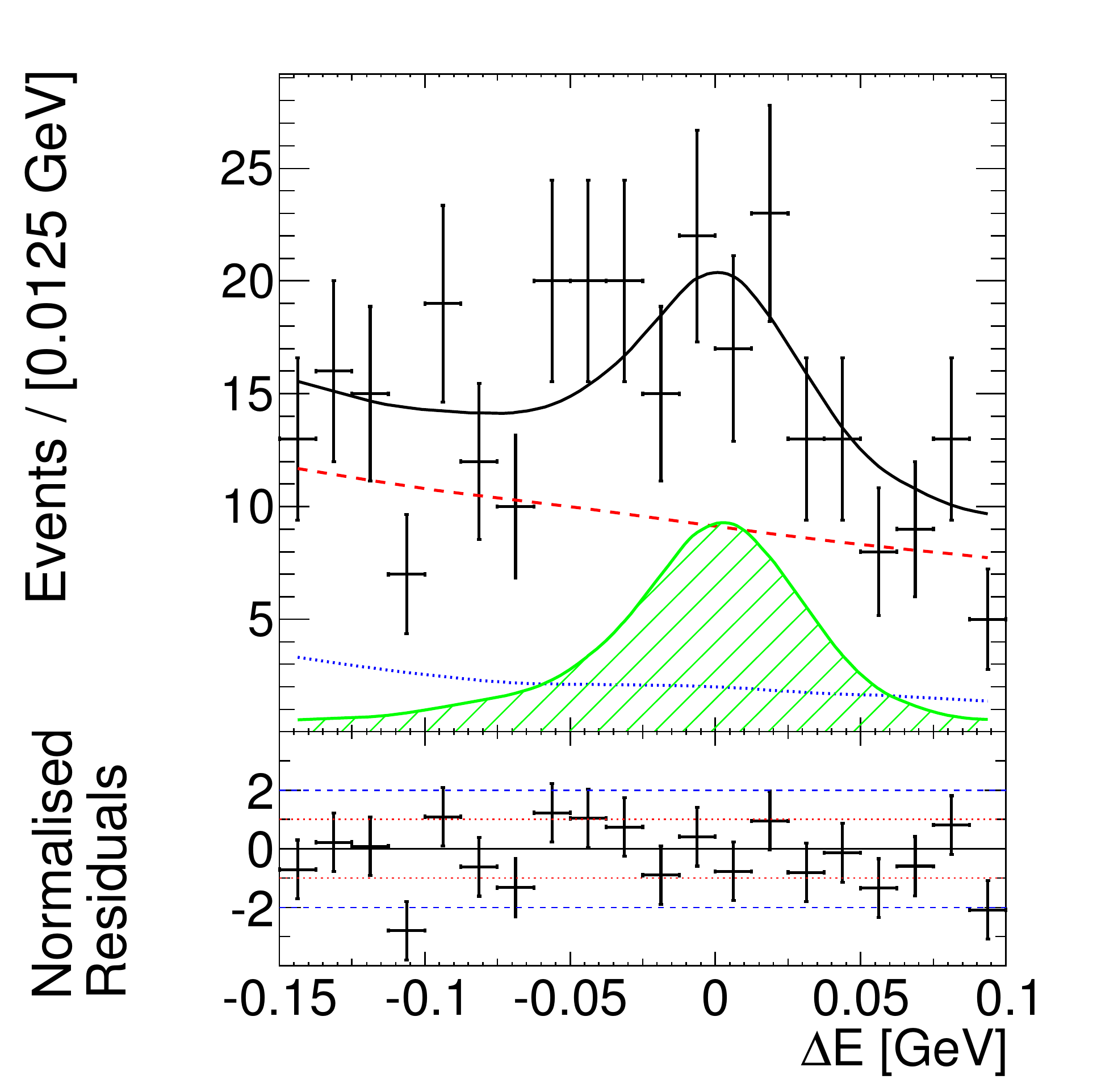}
  \includegraphics[width=0.3\textwidth]{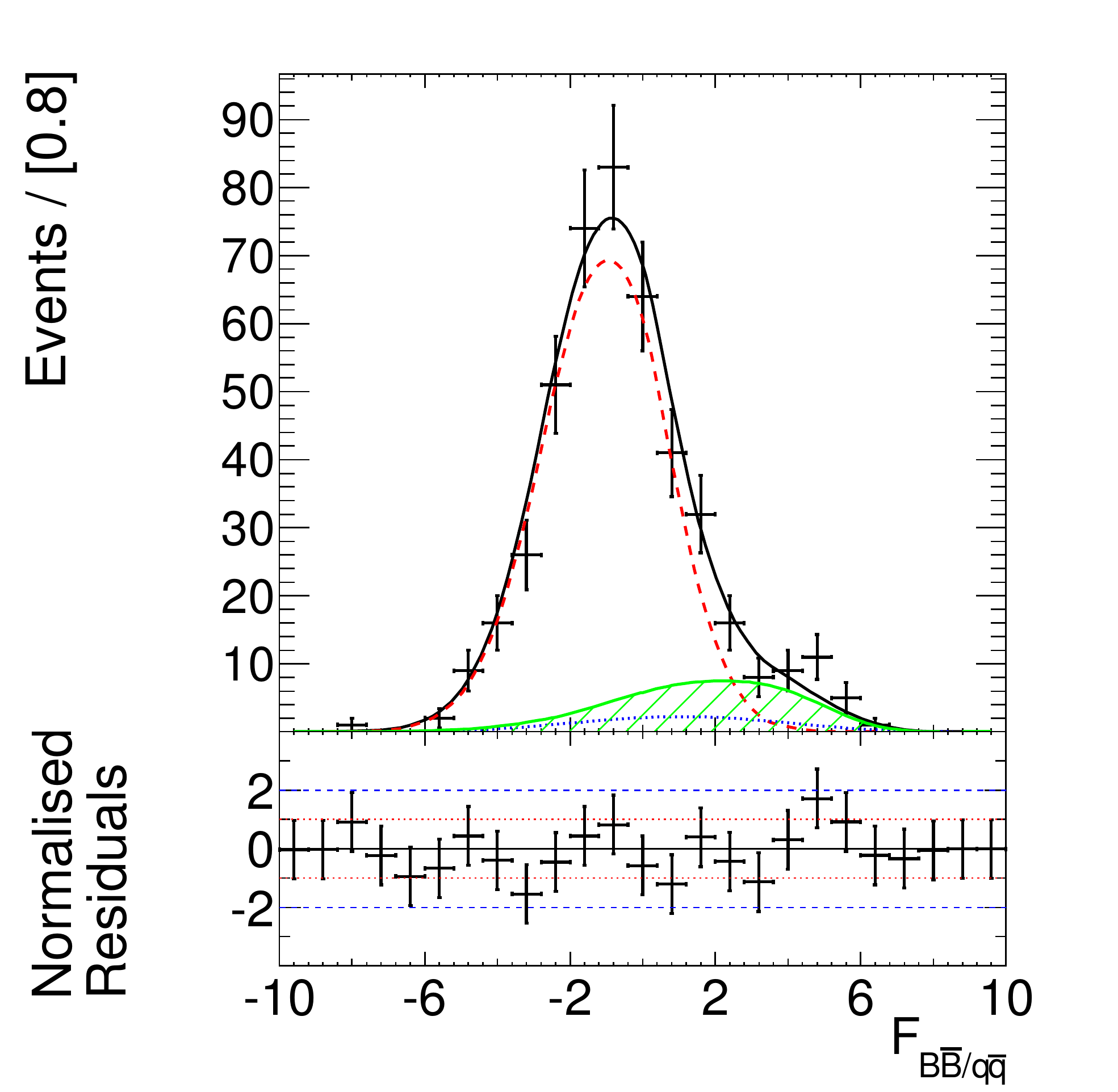}\\
  \includegraphics[width=0.3\textwidth]{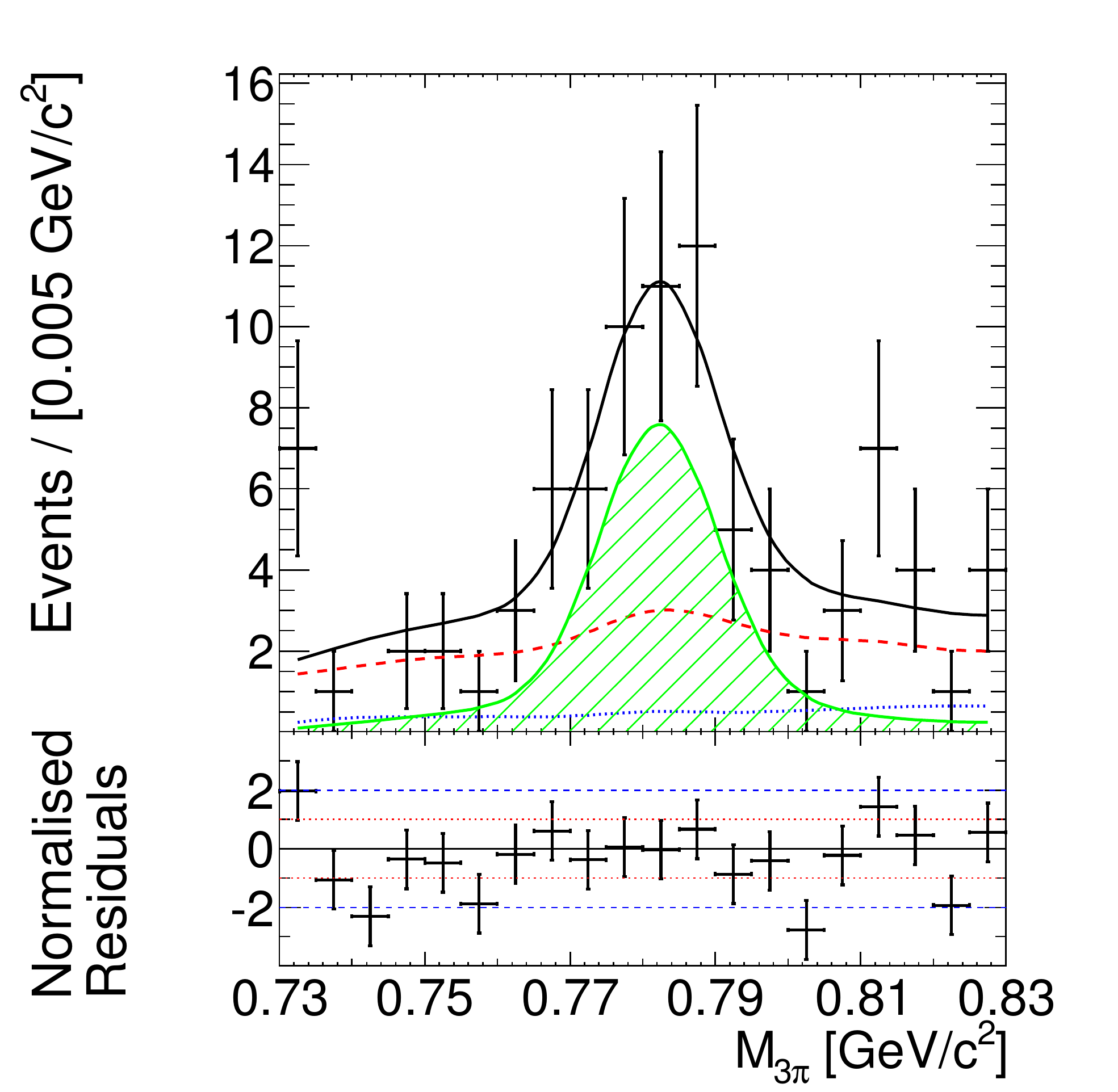}
  \includegraphics[width=0.3\textwidth]{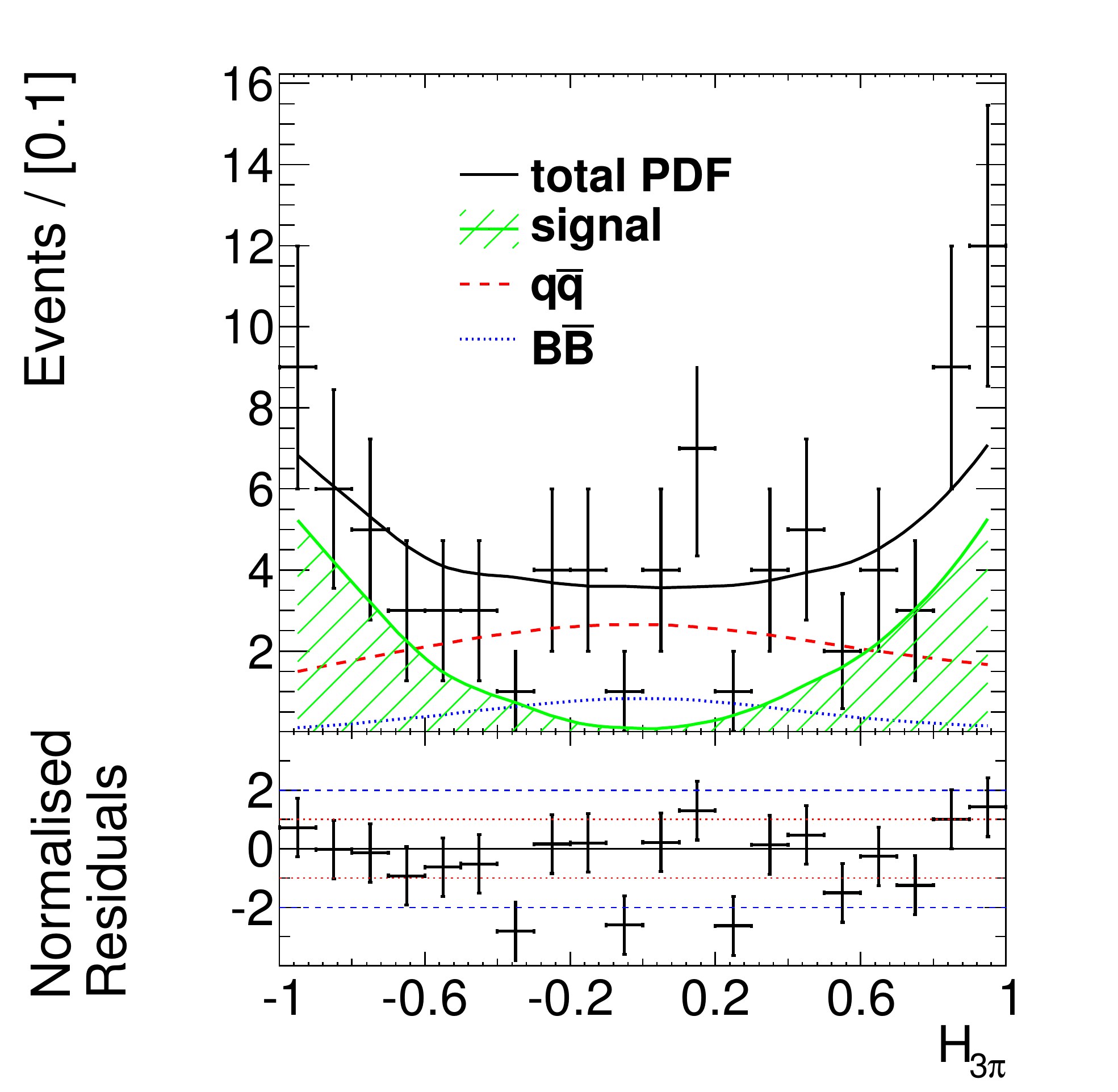}
  \includegraphics[width=0.3\textwidth]{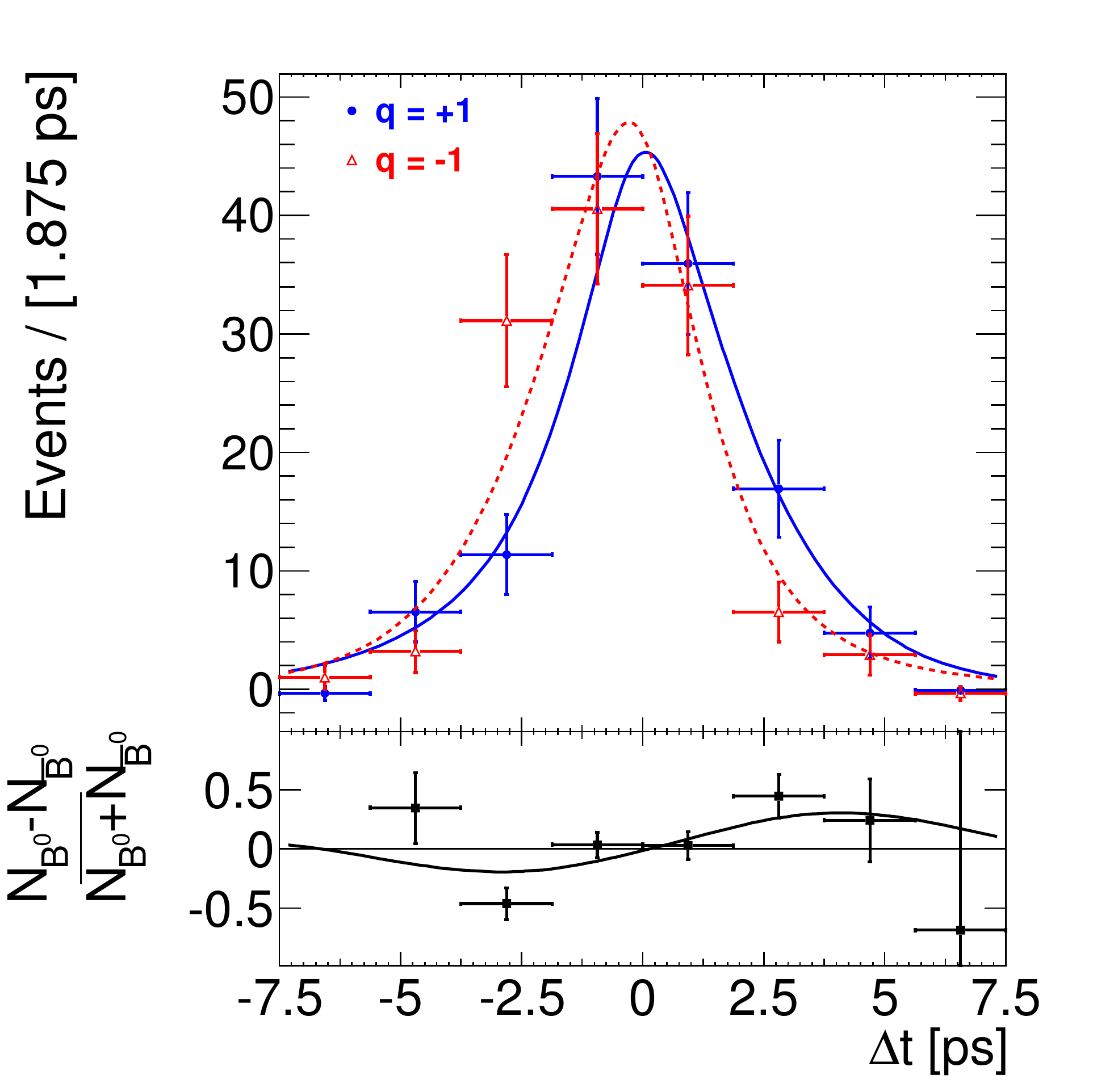}
  \caption{Projections of the data fit to the $B^{0}\rightarrow\omega K^{0}_{S}$ observables.}
  \label{wks_fit}
\end{figure}

We measure the branching fractions and the $CP$ parameters of the two decay modes in a seven-dimensional unbinned maximum likelihood fit to $M_{bc}$, $\Delta E$, the
mass $m_{3\pi}$ and helicity angle ${\cal H}_{3\pi}$ of the $\omega$ candidates, a Fisher discriminant ${\cal F}_{B\bar{B}/q\bar{q}}$, $\Delta t$ and the $B$ flavour 
$q$. The ﬁt is performed simultaneously to the neutral and charged modes, sharing common calibration factors. Following that, the shapes of the fit observables are fixed
and the ${\cal A}_{CP}$ parameter is obtained from two further fits to extract the number of $B^{+}$ and $B^{-}$ events. Projections of the the fit to the data are shown 
in Fig.~\ref{wks_fit}. We obtain
\begin{eqnarray}
  {\cal B}(B^{0}\rightarrow\omega K^{0}) &=& (4.5\pm0.4\,{\rm(stat)}\pm0.3\,{\rm(syst))\times 10^{-6}}, \nonumber \\
  {\cal B}(B^{-}\rightarrow\omega K^{-}) &=& (6.8\pm0.4\,{\rm(stat)}\pm0.4\,{\rm(syst))\times 10^{-6}}, \nonumber\\
  {\cal A}_{\omega K^{0}_{S}} &=& -0.36\pm0.19\,{\rm(stat)}\pm0.05\,{\rm(syst)}, \nonumber \\
  {\cal S}_{\omega K^{0}_{S}} &=& +0.91\pm0.32\,{\rm(stat)}\pm0.05\,{\rm(syst)}, \nonumber \\
  {\cal A}_{\omega K^{-}}&=& -0.03\pm0.04\,{\rm(stat)}\pm0.01\,{\rm(syst)}. \nonumber
\end{eqnarray}
For the branching fractions, the main contribution to the systematic uncertainties is the $\pi^{0}$ reconstruction efficiency. For ${\cal A}_{\omega K^{0}_{S}}$, the dominant 
systematic effect is the tag-side interference and for ${\cal S}_{\omega K^{0}_{S}}$ and ${\cal A}_{\omega K^{-}}$ - the $\Delta t$ resolution function. By performing a likelihood scan
in the ${\cal A}_{CP}-{\cal S}_{CP}$ plane, we find a deviation of $3.1\sigma$ from the $CP$ conservation hypothesis, which makes this result the first evidence for $CP$ violation in 
the $B^{0}\rightarrow\omega K^{0}_{S}$ decay mode. Apart from the ${\cal S}_{\omega K^{0}_{S}}$ result, this is the most precise measurement of the $B\rightarrow\omega K$ decay modes.
The results are mostly in agreement with the previous measurements of Belle~\cite{wk_belle_prev_1,wk_belle_prev_2} and BaBar~\cite{wk_babar_prev_1,wk_babar_prev_2}. The analysis has 
been published in the July 2014 issue of Physical Review D~\cite{wk_belle_final}. 

\section{Conclusion}
The results from the measurements on $b\rightarrow s q\bar{q}$ decays presented in these proceedings are consistent with the SM predictions and with the current world average on $\sin{2\phi_{1}}$
measured in $b\rightarrow c \bar{c} s$ transitions~\cite{sin2phi1_average}. The uncertainties on the parameter are still very large and so for new physics to be observed in these decay modes, 
more data is needed. This will be provided in the future by the Belle II and the LHCb experiments.


\end{document}